\documentclass{emulateapj}
\usepackage{apjfonts}

\begin{document}

\shortauthors{Zheng et al.}
\shorttitle{$I$-band Luminosity Function}

\title{M Dwarfs from {\it Hubble Space Telescope} Star Counts. V. The $I$-band
Luminosity Function\altaffilmark{1}}

\author{Zheng Zheng\altaffilmark{2}, Chris Flynn\altaffilmark{3},
        Andrew Gould\altaffilmark{2}, John N. Bahcall\altaffilmark{4},
        Samir Salim\altaffilmark{5}} 

\altaffiltext{1}{Based on observations with the NASA/ESA Hubble Space 
                 Telescope, obtained at the Space Telescope Science Institute, 
                 which is operated by the Association of Universities for 
                 Research in Astronomy, Inc. under NASA contract 
                 No. NAS5-26555.}
\altaffiltext{2}{Department of Astronomy, Ohio State University, Columbus, 
                 OH 43210; zhengz@astronomy.ohio-state.edu,  
                 gould@astronomy.ohio-state.edu.}  
\altaffiltext{3}{Tuorla Observatory, Turku University, 
                 V\"{a}is\"{a}l\"{a}ntie 20, FIN-21500, Piikki\"{o},
                 Finland; cflynn@astro.utu.fi.}
\altaffiltext{4}{Institute for Advanced Study, Princeton, NJ 08540;
                 jnb@ias.edu.}
\altaffiltext{5}{Department of Physics and Astronomy, University of California
                 at Los Angeles, Los Angeles, CA 90095; samir@astro.ucla.edu.} 

\begin{abstract}
We derive the disk $I$-band luminosity function from the Zheng et al. sample 
of $\sim 1400$ disk M dwarfs observed with the {\it Hubble Space Telescope}. 
We adopt a Galactic-height-dependent color-magnitude relation to account
for the metallicity gradient above the Galactic plane. The resultant $I$-band 
luminosity function peaks at $M_I \sim 9.5$ and drops sharply toward 
$M_I \sim 10.5$.
\end{abstract}

\section{Introduction}

The stellar luminosity function (LF) measures the number density of 
stars as a function of luminosity. It is traditionally expressed in the 
$V$-band. However, sometimes it is more convenient to use the $I$-band LF for 
planning and modeling observations. Here we present the $I$-band LF of disk M 
dwarfs from {\it Hubble Space Telescope} ({\it HST}) star counts.

\citet{Zheng01} studied a sample of about 1400 disk M dwarfs found in 
148 fields observed with the Wide Field Camera 2 (WFC2) on the {\it HST} and 
162 fields observed with pre-repair Planetary Camera 1 (PC1). The $V$-band LF
and the Galactic disk parameters were derived simultaneously using the method 
of maximum likelihood. If the luminosity of M dwarfs depended only on their 
color, as is the case for example for the solar neighborhood color-magnitude 
relation (CMR) determined by \citet{Reid91}, the conversion from the $V$-band 
LF to the $I$-band LF would be straightforward. However, since many stars in 
their sample lie far from the Galactic plane, \citet{Zheng01} introduced a 
more realistic CMR [CMR (2) in that paper], which makes a first-order 
correction for the metallicity effect: the luminosity then depends not only on 
the color but also on the height, $z$, above the Galactic plane. With this 
more complex CMR, the transformation from the $V$-band LF to the $I$-band LF 
is not trivial. 

In this paper, we apply the method of maximum likelihood to the same set of 
{\it HST} observations as in \citet{Zheng01} and incorporate the 
Galactic-height-dependent CMR to derive the $I$-band LF of M dwarfs. This
direct method avoids the uncertainty and the difficulty in the transformation 
from the $V$-band LF to the $I$-band LF. For details on the observations, the 
CMR, and the maximum likelihood method, we refer readers to \citet{Zheng01} and 
references therein. In \S~2, we present our results after a brief description 
of models.

\section{Models and Results}

The LF of disk M dwarfs is a function of the Galactic position ($R$, $z$). We 
parameterize it as the product of a vector of 12 absolute $I$-magnitude bins 
and a Galactic density law,
\begin{equation}
\Phi(i,z,R)=\Phi_i\nu(z)\exp\biggl(-{R-R_0 \over H}\biggr),
\end{equation}
where $\Phi_i$ is the solar neighborhood LF for the {\it i}th magnitude bin,
$R_0=8$ kpc is the Galactocentric distance of the Sun, and $H$ is the scale
length of the disk. The vertical density profile $\nu(z)$ is assumed to have
either a ``sech$^2$'' form,
\begin{equation}
\nu_s(z)=(1-\beta){\rm sech}^2 {z \over h_1} +\beta \exp 
\biggl(-{|z| \over h_2}\biggr),
\end{equation}
or a ``double-exponential'' form,
\begin{equation}
\nu_e(z)=(1-\beta) \exp\biggl(-{|z| \over h_1}\biggr) +\beta \exp 
\biggl(-{|z| \over h_2}\biggr).
\end{equation}
The magnitude bins are centered at $M_I=$ 6.5, 7.0, 7.5, 8.0, 8.5, 9.0,
9.5, 10.0, 10.5, 11.0, 12.0, and 13.5, respectively. The size of each bin
is 0.5 mag except the last two (1.5 mag). Given a choice of disk scale length 
$H$, the local LF, $\Phi_i$, and the three disk profile parameters ($h_1$, 
$h_2$, $\beta$) are found by maximizing the likelihood. The scale length
is determined by finding the maximum in the maximized likelihoods of
an ensemble of solutions using different values of $H$. 

As in \citet{Zheng01}, we select stars below $z=$ 2400 pc. The absolute 
$I$-magnitude of stars should satisfy $6.25<M_I<14.25$. The faint boundary 
is set to prevent contamination by spheroid giants. When translated into
color, the boundary we use here is slightly bluer than that in 
\citet{Zheng01}, but it is still acceptable since our results show virtually 
no change if we move the boundary to a higher absolute $I$-magnitude. 
Altogether, 1403 stars satisfy our selection criteria and they constitute 
our sample. This sample has 30 more stars than the one used by \citet{Zheng01}, 
which results primarily from the small difference in the blue boundary. 

We solve for the LF and disk profile parameters by maximizing the likelihood
and find that the resultant disk profile parameters, for either the sech$^2$
model or the double-exponential model, are almost identical to those derived 
in \citet{Zheng01}. This consistency of the results implies that the disk 
profile derived from the maximum likelihood method is robust. 
We therefore decide to fix the disk parameters at the values in 
\citet{Zheng01}, namely $h_1=$ 270 (156) pc, $h_2=$ 440 (439) pc, 
$\beta=$ 56.5\% (38.1\%), and $H=$ 2.75 kpc for the sech$^2$ 
(double-exponential) model. 

Similar to the the case in $V$-band, due to the lack of local stars in the 
sample, the best-fit LF for the sech$^2$ model and that for the 
double-exponential model have nearly the same shape but differ from each 
other in normalization, and the normalization difference is compensated by 
the vertical density profile. We adopt a linear combination of the LFs of
these two models as our final result, with the same combination coefficient 
derived by \citet{Zheng01} by matching the local normalization to the 
well-established luminous end of the solar neighborhood M-dwarf $V$-band LF.
 
The $I$-band LF from our analysis is listed in Table 1 and shown in Figure 1.
As expected, the overall shape of the $I$-band LF, with a peak at $M_I\sim 9.5$
and a sharp drop toward $M_I\sim 10.5$, mimics the $V$-band LF in 
\citet{Zheng01}.

\acknowledgments

We thank B. Paczy\'nski, whose request for an $I$-band LF was the genesis
of this paper. Work by Z.Z. and A.G. was supported in part by grant AST
02-01266 from the NSF. Z.Z. received additional support from a Presidential
Fellowship from the Graduate School of The Ohio State University. C. F. has 
been supported by the Academy of Finland through its support of the ANTARES 
program for space research.

\begin{deluxetable}{rr}
\tablecaption{The $I$-band Luminosity Function}
\tablehead{M$_I$ & $\phi$ \\
(mag) & (10$^{-3}$pc$^{-3}$$I$-mag$^{-1}$)}
\startdata
 6.5 &  4.5$\pm$1.1\\
 7.0 &  5.3$\pm$1.2\\
 7.5 &  6.0$\pm$1.2\\
 8.0 & 10.1$\pm$1.5\\
 8.5 & 13.0$\pm$1.8\\
 9.0 & 19.5$\pm$2.1\\
 9.5 & 23.1$\pm$2.2\\
10.0 & 14.8$\pm$1.8\\
10.5 &  7.6$\pm$1.5\\
11.0 &  8.4$\pm$1.6\\
12.0 &  3.4$\pm$0.6\\
13.5 &  3.3$\pm$0.9\\
\enddata
\end{deluxetable}

\begin{figure}[h]
\plotone{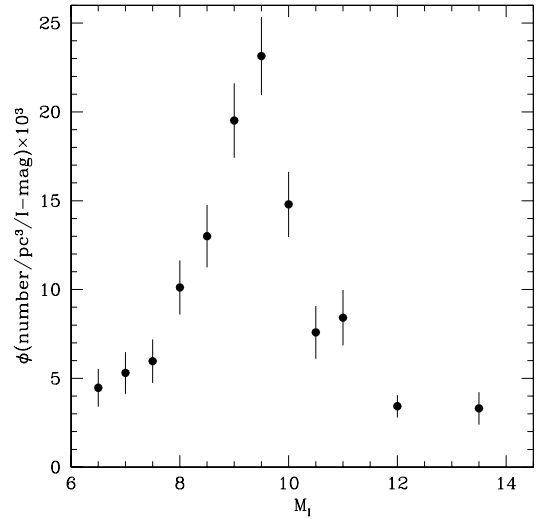}
\caption[]{The $I$-band luminosity function based on 1403 M dwarfs 
observed with {\it HST}.}
\end{figure}

\end{document}